\documentclass[aps,prb,reprint,showpacs,floatfix]{revtex4-1}
\usepackage{amsmath,amssymb,gensymb,graphicx,color,verbatim}

\begin{document}

\title{Composition-based phase stability model for multicomponent metal alloys}
\author{Jay C. Spendlove}
\altaffiliation[Present address: ]{Department of Physics and Astronomy, Brigham Young University, Provo, Utah 84602, USA}
\author{Bryan H. Fong}
\author{John H. Martin}
\author{Mark R. O'Masta}
\author{Andrew Pan}
\author{Tobias A. Schaedler}
\author{Eric B. Isaacs}
\email{ebisaacs@hrl.com}
\affiliation{HRL Laboratories, LLC, 3011 Malibu Canyon Road, Malibu, California 90265, USA}

\begin{abstract}
  The vastness of the space of possible multicomponent metal alloys is
  hoped to provide improved structural materials but also challenges
  traditional, low-throughput materials design efforts. Computational
  screening could narrow this search space if models for materials
  stability and desired properties exist that are sufficiently
  inexpensive and accurate to efficiently guide experiments. Towards
  this effort, here we develop a method to rapidly assess the
  thermodynamic stability of a metal alloy composition of arbitrary
  number of elements, stoichiometry, and temperature based on density
  functional theory (DFT) data. In our model, the Gibbs free energy of
  the solid solution contains binary enthalpy contributions and ideal
  configurational entropy, whereas only enthalpy is considered for
  intermetallic competing phases. Compared to a past model for
  predicting the formation of single-phase high-entropy alloys [Phys.
  Rev. X \textbf{5}, 011041 (2015)], our method is similarly
  inexpensive, since it assesses enthalpies based on existing DFT
  data, but less heuristic, more broadly applicable, and more accurate
  (70--75\%) compared to experiment.
\end{abstract}

\date{\today}
\maketitle

\section{Introduction}\label{sec:intro}

Multicomponent metal alloys, including those labeled ``high-entropy''
that contain several elements in non-dilute concentrations, constitute
a vast space of possible structural
materials.\cite{miracleCriticalReviewHigh2017,georgeHighentropyAlloys2019}
For example, there are approximately 3.5 billion possible compositions
that contain six out of 55 metallic elements in multiples of 10\%
(e.g.,
Cr$_{0.1}$Mn$_{0.2}$Fe$_{0.3}$Co$_{0.1}$Ni$_{0.2}$Os$_{0.1}$).\cite{widomFrequencyEstimateMulticomponent2017}
The vastness of this space, as compared to the smaller space of more
traditional alloys containing fewer elements, presents a significant
challenge to materials design efforts. Computational approaches to
help identify promising regions of composition space for a given
application, possibly in conjunction with the rapidly advancing
capabilities in high-throughput
experiments,\cite{mooreheadHighthroughputSynthesisMoNbTaW2020,meliaHighthroughputAdditiveManufacturing2020,vecchioHighthroughputRapidExperimental2021,miracleEmergingCapabilitiesHighThroughput2021,surHighThroughputDiscovery2023}
could accelerate materials discovery and
development.\cite{singhDesignHighstrengthRefractory2018,pollockEvolvingLandscapeAlloy2019,curtinProgressChallengesTheory2022,ouyangDesignRefractoryMultiprincipalelement2023,peiTheoryguidedDesignHighentropy2023}

In this work we focus on assessing whether a given hypothetical alloy
composition is realizable as a single-phase solid solution. Our target
is alloys that are thermodynamically stable (i.e., rather than
kinetically stable) given our interest in high-temperature
applications for which metastable materials may be less robust. We do
not here consider properties beyond stability, but we note that the
single-phase or multi-phase nature of an alloy can have a significant
impact on, for instance, mechanical properties and susceptibility to
corrosion.\cite{georgeHighEntropyAlloys2020,scullyControllingCorrosionResistance2020,zengMachineLearningAccelerated2023}

Taking inspiration from the Hume-Rothery rules, several researchers
have developed empirical-type heuristic rules to predict the formation
of single-phase multicomponent alloys based on avoiding large
differences in atomic size, electronegativity, and valence electron
concentration among the constituent
elements.\cite{guoMoreEntropyHighentropy2013,zhangGuidelinesPredictingPhase2014,gaoThermodynamicsConcentratedSolid2017}
The CALPHAD (Calculation of Phase Diagrams) method and machine
learning approaches, which provide flexible frameworks for phase
stability predictions but rely heavily on pre-existing experimental
training data, are also in use for such
purposes.\cite{gaoThermodynamicsConcentratedSolid2017,saalEquilibriumHighEntropy2018,liMachinelearningModelPredicting2019,peiMachinelearningInformedPrediction2020,wangInsightsPhaseFormation2022,vazquezDeepNeuralNetwork2023}
First-principles electronic structure calculations of alloy free
energy\cite{fengFirstprinciplesPredictionHighentropyalloy2017,ledererSearchHighEntropy2018,ikedaInitioPhaseStabilities2019,niuMulticellMonteCarlo2019}
in principle could provide a rigorous, accurate, and broadly
applicable alternative model that avoids errors relating to limited
experimental training data, which is particularly important for
application to unexplored parts of composition space.

Along these lines, Troparevsky \textit{et al.} developed a model to
predict the formation of a single-phase solid solution purely based on
density functional theory (DFT)
energetics.\cite{troparevskyCriteriaPredictingFormation2015} They
reasoned that a single-phase solid solution will not form if there
exists, for at least one pair of constituent elements, either (1) one
or more highly stable (i.e., very negative formation enthalpy) binary
intermetallic phases or (2) all highly unstable (i.e., very positive
formation enthalpy) binary intermetallic phases. A multiphase alloy is
expected for the first case due to precipitation of such
intermetallic(s) and for the second case related to a tendency for
elemental phase separation. The criteria were assessed by looking up,
for each pair of constituent elements, the lowest binary intermetallic
formation enthalpy (for an ordered structure of any stoichiometry)
present in an existing DFT database. The free energy contribution
corresponding to the ideal configurational entropy of a five-element
equiatomic composition and a typical experimental annealing
temperature was used to set the lower formation enthalpy limit,
whereas the upper limit was chosen empirically.

Here, we develop a thermodynamic model that is based on the same
well-reasoned physical intuition of Troparevsky \textit{et al.} and
retains its computational affordability, but has several key
advantages. Since we explicitly model the stoichiometry-dependent
Gibbs free energy and determine ground states via the convex hull
construction, our model (1) is a proper stoichiometry-dependent model
that takes into account, for example, the location of competing
intermetallic phases in composition space, (2) has no heuristic and/or
empirical formation enthalpy bounds, (3) provides more complete
information, such as the complete set of phases expected to form for a
multiphase alloy, and (4) is applicable to both equiatomic
compositions, the focus of the model of Troparevsky \textit{et al.},
and non-equiatomic compositions. Our model achieves an accuracy of
70--75\% when validated on high-entropy alloy experiments and provides
a platform upon which additional physical effects (e.g., short-range
order, vibrational entropy, defect configurational entropy) could be
incorporated for improved accuracy.

\section{Thermodynamic Model}\label{sec:model}

We consider a single solid solution phase and a finite number of
ordered intermetallic competing phases that are treated as line
compounds. The intermetallic phases are among the approximately one
million materials in the Open Quantum Materials Database
(OQMD),\cite{kirklinOpenQuantumMaterials2015,shenReflectionsOneMillion2022}
which contains DFT calculations of both experimentally known phases
from the Inorganic Crystal Structure Database (ICSD) and hypothetical
ones derived from decoration of crystal prototypes. Our model is a
composition-only model in that we consider a single solid solution
phase of unspecified crystal structure at each composition.

Our model for the Gibbs (formation) free energy for the solid solution
$\Delta G^{\mathrm{alloy}}$ at temperature $T$ consists of an enthalpy
term $\Delta H^{\mathrm{alloy}}$ and a contribution from the entropy
$\Delta S^{\mathrm{alloy}}$: \begin{equation}
  \Delta G^{\mathrm{alloy}} = \Delta H^{\mathrm{alloy}} - T \Delta S^{\mathrm{alloy}}.
  \label{eq:alloygibbs}
\end{equation}
Since we only consider dense solid phases at ambient pressure in this
work, the pressure--volume term in the enthalpy is negligible and we
take all formation enthalpies to equal formation energies.


The entropy is approximated as the ideal configurational entropy
$\Delta S^{\mathrm{alloy}} = - k_B \sum_i x_i \ln(x_i),$ where $x_i$
is the atomic percentage of each element $i$ in the composition
(normalized with $\sum_i x_i = 1$) and $k_B$ is the Boltzmann
constant. The alloy entropy thus has no temperature dependence in our
model. Other entropy contributions such as non-ideal configurational
entropy and vibrational entropy are not included in this work to avoid
a more expensive and complex model.

The alloy enthalpy, similarly treated as independent of temperature,
is derived from (zero-temperature) DFT energetics. To evaluate the
solid solution enthalpy term, we first construct the enthalpy of a
binary solid solution $\Delta H_{\alpha,\beta}^{\mathrm{alloy}}$ (for
distinct elements $\alpha$ and $\beta$). In principle, one can
estimate the enthalpy of the solid solution for a particular lattice
by averaging the enthalpies of many different ordered structures on
that lattice, to the extent the enthalpy is a well-behaved function of
local atomic configurations and the ordered structures appropriately
sample configurations contained in the solid solution. As an
illustration of this, the solid solution formation energy obtained via
quasirandom structure calculations is commonly seen to take on an
average of the formation energies of ordered structures considered
when generating a cluster expansion (see, for example, Ref.
\citenum{huaFirstprinciplesStudyVibrational2018}). Taking inspiration
from this property, we determine
$\Delta H_{\alpha,\beta}^{\mathrm{alloy}}$ based on the formation
energies of intermetallic phases in the OQMD, which includes various
binary crystal structures that are orderings on the fcc (e.g.,
D0$_{22}$, L1$_0$, L1$_1$, L1$_2$), bcc (e.g., B2, D0$_3$, B$_{19}$),
hcp (e.g., D0$_{19}$), and other (e.g., B1, B3, B4, B$_{\mathrm{h}}$)
lattices for every $\alpha$--$\beta$ pair, in addition to experimental
crystal structures from the ICSD. This approach is similar in spirit
to the small set of ordered structures (SSOS)
method,\cite{jiangEfficientInitioModeling2016} although we do not
consider any weighting of different ordered structures based on their
local structural correlations. Since intermetallic phases
corresponding to multiple underlying lattices are present, we obtain a
general (i.e., not associated with a specific lattice) enthalpy
tendency suitable for our composition-only model.

To obtain an average, we fit the ordered structure formation energies
to the polynomial form
\begin{equation}
  \Delta H_{\alpha,\beta}^{\mathrm{alloy}}(x_{\alpha}) = A_{\alpha,\beta}x_{\alpha}(1-x_{\alpha}),
  \label{eq:binaryenthalpy}
\end{equation} where $A$ is a scalar fitting parameter encapsulating
the enthalpy behavior for each elemental pair. The fitting form in Eq. \ref{eq:binaryenthalpy} is the zeroth-order
Redlich-Kister
polynomial\cite{redlichAlgebraicRepresentationThermodynamic1948} and
corresponds to the enthalpy term in a regular solution model. However,
since we are not fitting the formation energies of phases that all
correspond the same parent lattice, we can view the $x(1-x)$ form as
essentially a simple fitting function to capture the overall binary
enthalpy trend. To avoid disruption of the fit by any outlier phases
(e.g., stemming from structures that are extremely high energy for the
composition), we ignore any phases whose formation energy is beyond
the mean value, for the full binary space, by more than 1.5 times the
standard deviation.

We use the fitted binary formation enthalpies to estimate the
formation enthalpy of the multicomponent solid solution (containing
$N$ elements)
via the linear interpolation equation
\begin{equation}
  \Delta H^{\mathrm{alloy}} = \sum_{\alpha,\ \beta > \alpha} w_{\alpha,\beta} \Delta H_{\alpha,\beta}^{\mathrm{alloy}}\!\!\left(x=\frac{x_{\alpha}}{x_{\alpha}+x_{\beta}}\right),
  \label{eq:enthalpyinterpolation}
\end{equation}
where the sum is over distinct elemental pairs. As in the Kohler
method,\cite{kohlerZurBerechnungThermodynamischen1960} here the
stoichiometry of the relevant binary composition is determined simply
by the relative amount of the two elements, i.e.,
$x_{\alpha}/(x_{\alpha}+x_{\beta})$. The weighting factor
$w_{\alpha,\beta} = (x_{\alpha} + x_{\beta})/(N-1)$ is chosen to
preserve mass balance. As a simple example, in this method the
formation enthalpy of NbTaW is equal to an average of
$\Delta H_{\mathrm{Nb,Ta}}^{\mathrm{alloy}}(x=1/2)$,
$\Delta H_{\mathrm{Ta,W}}^{\mathrm{alloy}}(x=1/2)$, and
$\Delta H_{\mathrm{W,Nb}}^{\mathrm{alloy}}(x=1/2)$. We note that
alternative interpolation schemes such as Muggianu's approach
exist,\cite{muggianuEnthalpiesFormationAlliages1975,hillertEmpiricalMethodsPredicting1980,hillert2007phase,meredigCombinatorialScreeningNew2014,kattnerCALPHADMETHODITS2016}
but they are not explored in this work given we expect the binary
enthalpy contributions themselves to be a more important source of
uncertainty.

We consider two enthalpy models for the multicomponent alloy. In the
first, we only include a contribution from
$\Delta H_{\alpha,\beta}^{\mathrm{alloy}}$ if all intermetallic phases
in the $\alpha$--$\beta$ binary space are unstable, i.e.,
$w_{\alpha,\beta}$ is set to zero if there are any stable (negative
formation energy) intermetallics in the space. This enthalpy model
(labeled ``only positive enthalpy contributions'') is in the spirit of
the work of Troparevsky \textit{et al.} since in their model, with
respect to the upper formation energy limit, it is only the presence
of an elemental pair with all sufficiently unstable formation energies
--- regardless of possible stabilizing enthalpic contributions from
other elemental pairs --- that leads to the prediction of an unstable
solid solution. By contrast, in our second enthalpy model (labeled
``positive and negative enthalpy contributions''), we simply include
contributions from all elemental pairs.

Finally, for the Gibbs free energy of the intermetallic competing
phases $\Delta G^{\mathrm{intermetallic}}$, we only include the
formation enthalpy part (obtained directly from the OQMD), i.e., they
have no entropy. Configurational entropy associated with
defects/off-stoichiometry and vibrational entropy are not included to
avoid the need for expensive calculations. We note that since the
Gibbs free energy of the intermetallic competing phases therefore has
no temperature dependence, only intermetallics that appear on the
zero-temperature convex hull have the possibility to appear on the
Gibbs free energy convex hull (at any temperature) in our model.

\begin{figure}[htbp]
  \begin{center}
    \includegraphics[width=1.0\linewidth]{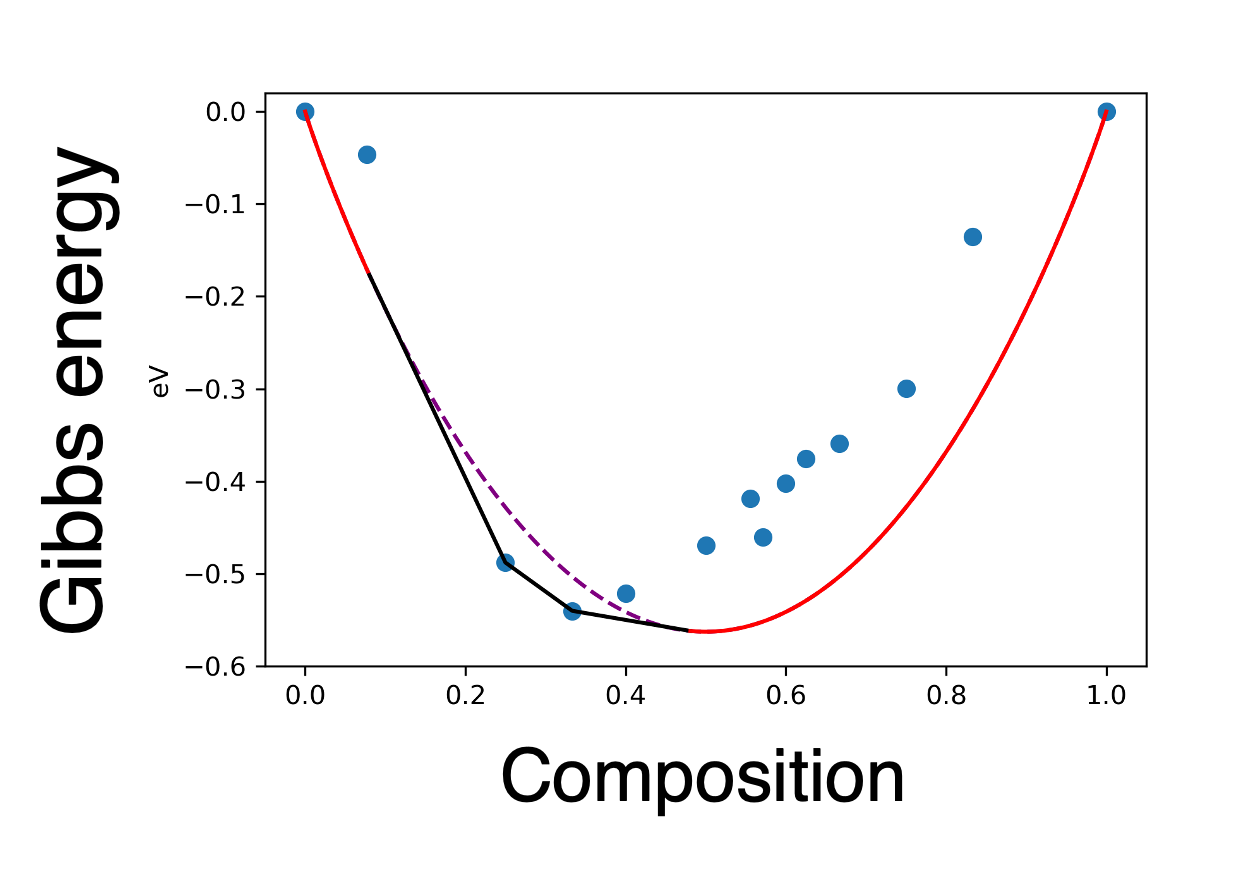}
  \end{center}
  \caption{Schematic of the convex hull of Gibbs free energy in our
    model for a binary case. The blue points correspond to
    intermetallic competing phases (and the pure elements at the
    compositions $x=0$ and $x=1$) and the dashed purple line
    corresponds to the solid solution. The convex hull consists of (1)
    regions with a single-phase solid solution ground state (red
    curves) and (2) multiphase regions (black lines).}
  \label{fig:convex_hull_schematic}
\end{figure}

Figure \ref{fig:convex_hull_schematic} shows a schematic example of
the Gibbs free energy behavior within our model. The solid lines are
the convex hull of Gibbs free energy versus composition (for a
particular temperature), which is determined from the solid solution
Gibbs free energy $\Delta G^{\mathrm{alloy}}$ (dashed line) and those
of the intermetallic competing phases
$\Delta G^{\mathrm{intermetallic}}$ (points). The black segments
correspond to multiphase cases in which at least one intermetallic
phase is contained in the thermodynamic ground state. In contrast, the
red parts of the convex hull (whose limits are determined by, e.g.,
the common tangent construction) are regions in which the ground state
is a single-phase solid solution. Only a binary case is shown for
simplicity but the behavior for ternary and higher cases is analogous.
We note that in the actual implementation of our thermodynamic model
for the general multicomponent case, we employ a discretized
stoichiometry grid for the solid solution phase to enable use of the
existing convex hull construction capabilities in the OQMD based on
the quick hull algorithm in
\texttt{Qhull},\cite{barberQuickhullAlgorithmConvex1996} as opposed to
common tangent construction.

The schematic in Fig. \ref{fig:convex_hull_schematic} is drawn to
emphasize the possibility of, considering solely the intermetallic
phases, a convex hull that is appreciably asymmetric, i.e., not
invariant to permutation of different elements. Given our model is
capable of representing such an asymmetry via full inclusion of the
intermetallic competing phases, it can capture the resulting asymmetry
of the predicted region(s) of single-phase solid solution. This
contrasts with approaches that only consider the value of an
intermetallic formation enthalpy and not its corresponding
stoichiometry.

\section{Results and Discussion}\label{sec:results}

\begin{figure}[htbp]
  \begin{center}
    \includegraphics[width=1.0\linewidth]{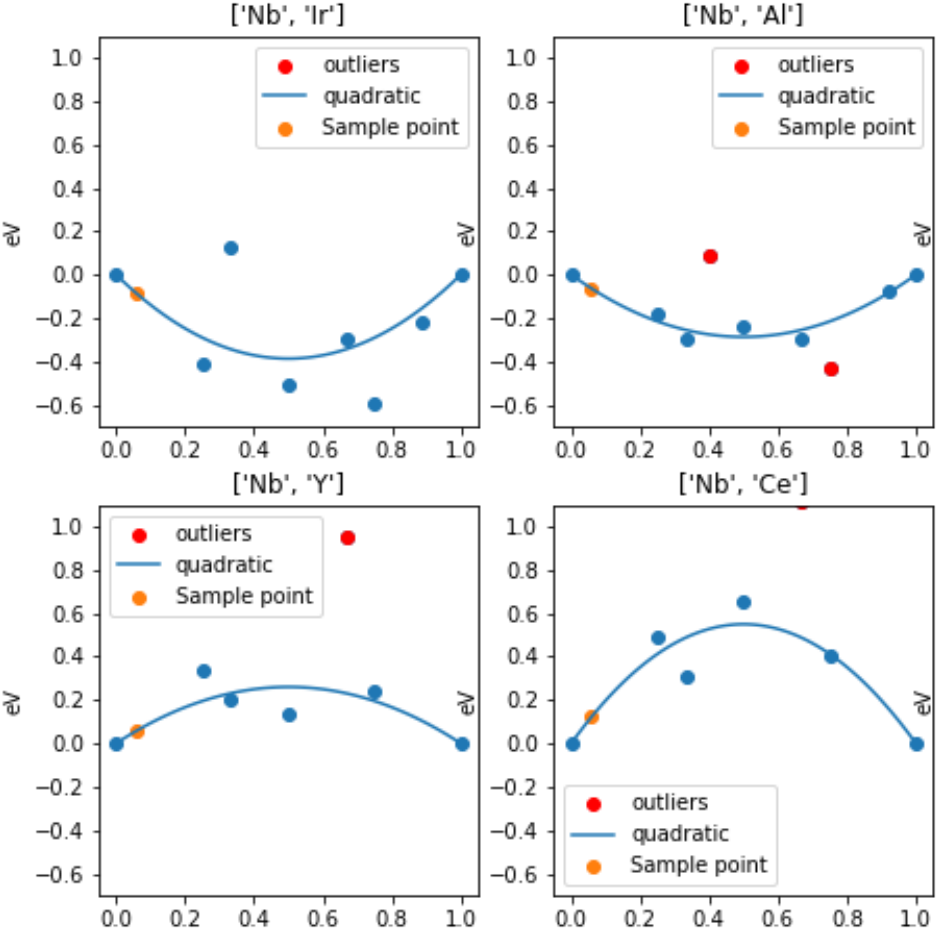}
  \end{center}
  \caption{Binary formation enthalpy
    $\Delta H_{\alpha,\beta}^{\mathrm{alloy}}(x_{\alpha}) =
    A_{\alpha,\beta}x_{\alpha}(1-x_{\alpha})$ fit (blue line) to
    formation energies of ordered intermetallic structures (blue
    points) in the OQMD for four elemental pairs: Nb--Ir, Nb--Al,
    Nb--Y, and Nb--Ce. The abscissa is $1-x_{\mathrm{Nb}}$. Red points
    are considered outliers and not included in the fit. The orange
    point indicates the region of enthalpy values that contributes to
    the multicomponent alloy enthalpy $\Delta H^{\mathrm{alloy}}$ for
    a Nb-rich alloy.}
  \label{fig:binaryfitting}
\end{figure}

Figure \ref{fig:binaryfitting} shows the fitting of
$\Delta H_{\alpha,\beta}^{\mathrm{alloy}}(x_{\alpha})$ for four
example elemental pairs: Nb with Ir, Al, Y, and Ce. Nb--Ir and Nb--Al
generally have negative (stable) formation energies leading to a
negative (favoring mixing) enthalpy contribution, in contrast to the
phase separating tendency of Nb--Y and Nb--Ce. We observe that the
fits capture the overall enthalpy trends reasonably well. The
symmetric (about $x_{\alpha}=1/2$) nature of the zeroth-order
Redlich-Kister polynomial does not appear problematic given the
overall formation energy behavior (distinct from the enthalpy convex
hull discussed in Sec. \ref{sec:model}) is not especially asymmetric.
Additional examples of fits of
$\Delta H_{\alpha,\beta}^{\mathrm{alloy}}(x_{\alpha})$ are shown in
the Supplemental Material.

\begin{figure}[htbp]
  \begin{center}
    \includegraphics[width=1.0\linewidth]{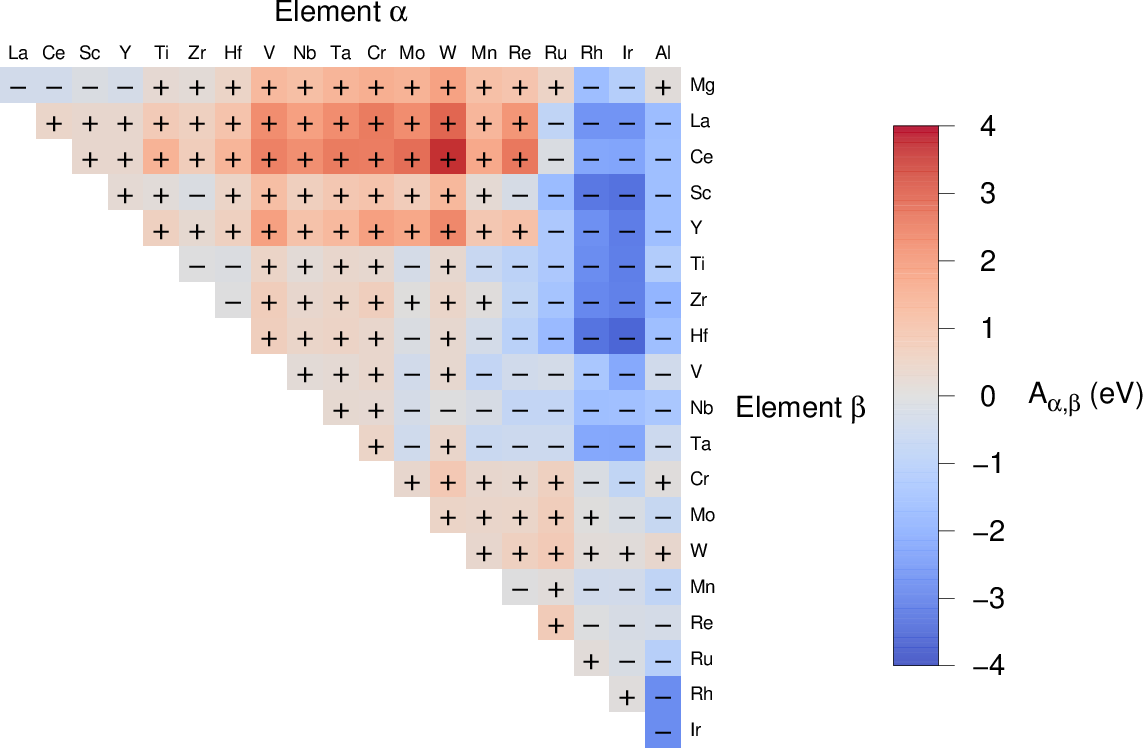}
  \end{center}
  \caption{$A_{\alpha,\beta}$ parameters representing the enthalpy
    behavior for a given pair of metal elements. Only the top-right of
    the table is shown as the bottom-left contains redundant
    information ($A_{\alpha,\beta}= A_{\beta,\alpha}$).}
  \label{fig:our_rainbow_matrix}
\end{figure}

A set of the fit $A_{\alpha,\beta}$ parameters considered in this work
are displayed in Fig. \ref{fig:our_rainbow_matrix}. Like the table of
the single most stable intermetallic formation energy contained in
Fig. 1 of Ref. \citenum{troparevskyCriteriaPredictingFormation2015},
this matrix is determined via a large number of DFT calculations
(derived from the OQMD in this work). In the present case, the matrix
quantifies the overall binary miscibility trends for the full
stoichiometry range.

For the set of elements shown in Fig. \ref{fig:our_rainbow_matrix}, we
observe a region of strong phase separation for elemental pairs
including one of the lanthanides (La or Ce) with Re or an element from
group V or VI. In contrast, strongly negative enthalpy contributions
are found for elemental pairs with (1) Rh or Ir and (2) one of several
rare-earth elements (La, Ce, Sc, Y), refractory elements (Ti, Zr, Hf,
Ta), or Al. We note that the presence of very stable binary
intermetallic competing phases, which may prevent the formation of a
single-phase solid solution, contributes to the negative binary
formation enthalpy contribution. Some similar trends can be observed
in the corresponding table from Troparevsky \textit{et al.}, but there
are significant quantitative differences since Fig.
\ref{fig:our_rainbow_matrix} represents the overall enthalpy trend as
opposed to the single lowest formation energy. Similarly, the
corresponding $A_{\alpha,\beta}$ values obtained via the semiempirical
Miedema model (commonly used to estimate multicomponent enthalpy; see,
e.g., Refs.
\citenum{guoMoreEntropyHighentropy2013,kingPredictingFormationStability2016,liMachinelearningModelPredicting2019})
have similar broad trends but are quantitatively very different, as
shown in the Supplemental Material.



To validate our thermodynamic model, we assess how accurately it
predicts whether a given multicomponent alloy composition corresponds
to a single-phase solid solution (classified here as ``positive'') as
opposed to a combination of multiple phases (classified as
``negative'') in experiment. The positive (negative) case corresponds
to the red (black) region in Fig. \ref{fig:convex_hull_schematic}. In
attempting to compare to experiment, several challenges include (1)
uncertainty of whether an alloy is truly thermodynamically (rather
than kinetically) stable, (2) dependence on synthesis method and
processing conditions (not generally captured in our model), (3)
experimental limitations in detecting small amounts of secondary
phases, and (4) limited amounts of experimental data.

We apply our model to 90 cases selected from a database of
experimental measurements of high-entropy
alloys.\cite{borgExpandedDatasetMechanical2020} We only consider
annealed samples and apply our thermodynamic model at the annealing
temperature, with the assumption that the annealing treatment is
sufficient to achieve thermodynamic equilibrium and therefore the
observed single-phase or multiphase nature of the alloy is dictated by
the thermodynamic phase stability at that temperature. We note that
the presence of multiple \textit{distinct} solid solution phases
(e.g., with different lattices or rich in different elements) can
occur in experiment but is not observed in our model due in part to
its composition-only nature, which represents one source of error.

For comparison, we also test the model of Troparevsky \textit{et al.}
While their model used DFT data primarily from the AFLOW
database,\cite{troparevskyCriteriaPredictingFormation2015} here we use
data in the OQMD to enable a fair comparison to our thermodynamic
model. We do not find significant differences in predictions stemming
from using the OQMD instead of AFLOW, which can be understood in part
by the good agreement between the OQMD and AFLOW for intermetallic
formation energies (in particular, median absolute difference of
2.7\%).\cite{hegdeQuantifyingUncertaintyHighthroughput2023} Additional
details of the validation are included in the Supplemental Material.


\begin{table}
  \begin{center}
    \begin{tabular}{|l|r|r|r|r|r|}\hline
      Model & TP & TN & FP & FN & Accuracy\\
      \hline
      (a) This work, only \(+\) & 29 & 38 & 4 & 19 & 74\%\\
      (b) This work, \(+\) and \(-\) & 34 & 28 & 14 & 14 & 69\%\\
      (c) Troparevsky et al. & 16 & 36 & 6 & 32 & 58\%\\\hline
    \end{tabular}
    \caption{Truth table for (a) our thermodynamic model with only
      positive binary enthalpy contributions, (b) our thermodynamic
      model with both positive and negative binary enthalpy
      contributions, and (c) the model of Troparevsky \textit{et al.}
      We (arbitrarily) assign positive (P) to single-phase and
      negative (N) to multiphase. True (T) corresponds to correct
      (compared to experiment) prediction of the model and false (F)
      corresponds to incorrect prediction. Accuracy is the percentage
      of predictions that are true.}
    \label{tab:truth}
  \end{center}
\end{table}

Table \ref{tab:truth} shows the validation results in the form of a
truth table. For our thermodynamic model, we test both the version
that only includes $\Delta H_{\alpha,\beta}^{\mathrm{alloy}}$ for
elemental pairs if all intermetallic phases in the $\alpha$--$\beta$
binary space are unstable and the version with enthalpy contributions
from all elemental pairs, as described in Sec. \ref{sec:model}. The
best model in terms of accuracy is the model with only positive
enthalpy contributions, which has an accuracy of 74\%. If we include
both positive and negative enthalpy contributions, the accuracy is
slightly lower at 69\%. However, given including both positive and
negative enthalpy contributions provides a better balance of false
positives and false negatives, we consider this model as the most
robust. The model of Troparevsky \textit{et al.} achieves an
appreciably lower accuracy of 58\% and it has the most unbalanced
distribution of false positive and false negatives, with over five
times as many false negatives as false positives.

The improved performance of our thermodynamic model indicates the
utility of a rigorous, non-heuristic approach to phase stability
predictions for multicomponent alloys. Furthermore, the nonempirical
nature of the method offers the potential for excellent
transferability given the absence of any parameters fit to experiments
on a specific chemistry. While the increased accuracy is desirable, we
view the broadened applicability (to arbitrary temperature and
stoichiometry) and more complete information (e.g., full set of
competing phases) to be a key benefit of our model. For instance, our
approach can be straightforwardly employed to help search for alloy
compositions amenable to precipitation hardening, in addition to
single-phase alloys.

With correct predictions observed approximately three out of four
times, the thermodynamic model developed in this work appears to be
moderately accurate but far from perfect, which is understandable
given (1) the various neglected physical effects and significant
approximations involved in the model construction (in part to ensure
low computational cost) and (2) uncertainty in the experimental
results. Therefore, we view the developed model as most applicable as
part of broad, high-throughput searches (e.g., as an initial
downselect), rather than for making highly quantitatively accurate
predictions for a small number of compositions of interest. With the
ability to utilize standard information existing in DFT databases, our
approach can enable the acceleration of discovery and development
within the vast space of multicomponent alloys.

\section{Conclusions}

We develop a first-principles phase stability model for multicomponent
metal alloys that considers an alloy composition of unspecified
crystal structure. Taking inspiration from the more heuristic model of
Troparevsky \textit{et al.}, we build a rigorous model by explicitly
constructing the Gibbs free energy of the solid solution phase, with
binary enthalpy contributions and ideal configurational entropy, and
that of intermetallic competing phases, for which only enthalpy is
considered. Our model captures the tendencies for intermetallic
formation and elemental phase separation that can prevent formation of
a single-phase solid solution. Since our model is computationally
inexpensive (only requiring energies from DFT calculations that
already exist in a database) and is reasonably accurate at classifying
an alloy as single-phase or multiphase, it can aid high-throughput
efforts for the discovery and development of multicomponent alloys.

\begin{acknowledgments}
  Funding was provided by HRL Laboratories, LLC. We acknowledge
  Cameron Cook, Greg Rutkowski, Brennan Yahata, Marc Dvorak, Yuksel
  Yabansu, Jake Hundley, Geoff McKnight, Dick Cheng, and Andy Detor
  for valuable discussions.
\end{acknowledgments}

\bibliography{stability_model}

\end{document}


\title{Supplemental Material for ``Composition-based phase stability model for multicomponent metal alloys''}
\author{Jay C. Spendlove}
\altaffiliation[Present address: ]{Department of Physics and Astronomy, Brigham Young University, Provo, Utah 84602, USA}
\author{Bryan H. Fong}
\author{John H. Martin}
\author{Mark R. O'Masta}
\author{Andrew Pan}
\author{Tobias A. Schaedler}
\author{Eric B. Isaacs}
\email{ebisaacs@hrl.com}
\affiliation{HRL Laboratories, LLC, 3011 Malibu Canyon Road, Malibu, California 90265, USA}

\date{\today}
\maketitle

\section{Additional examples of binary enthalpy fits}

\begin{figure}[h]
  \begin{center}
    \includegraphics[width=0.5\linewidth]{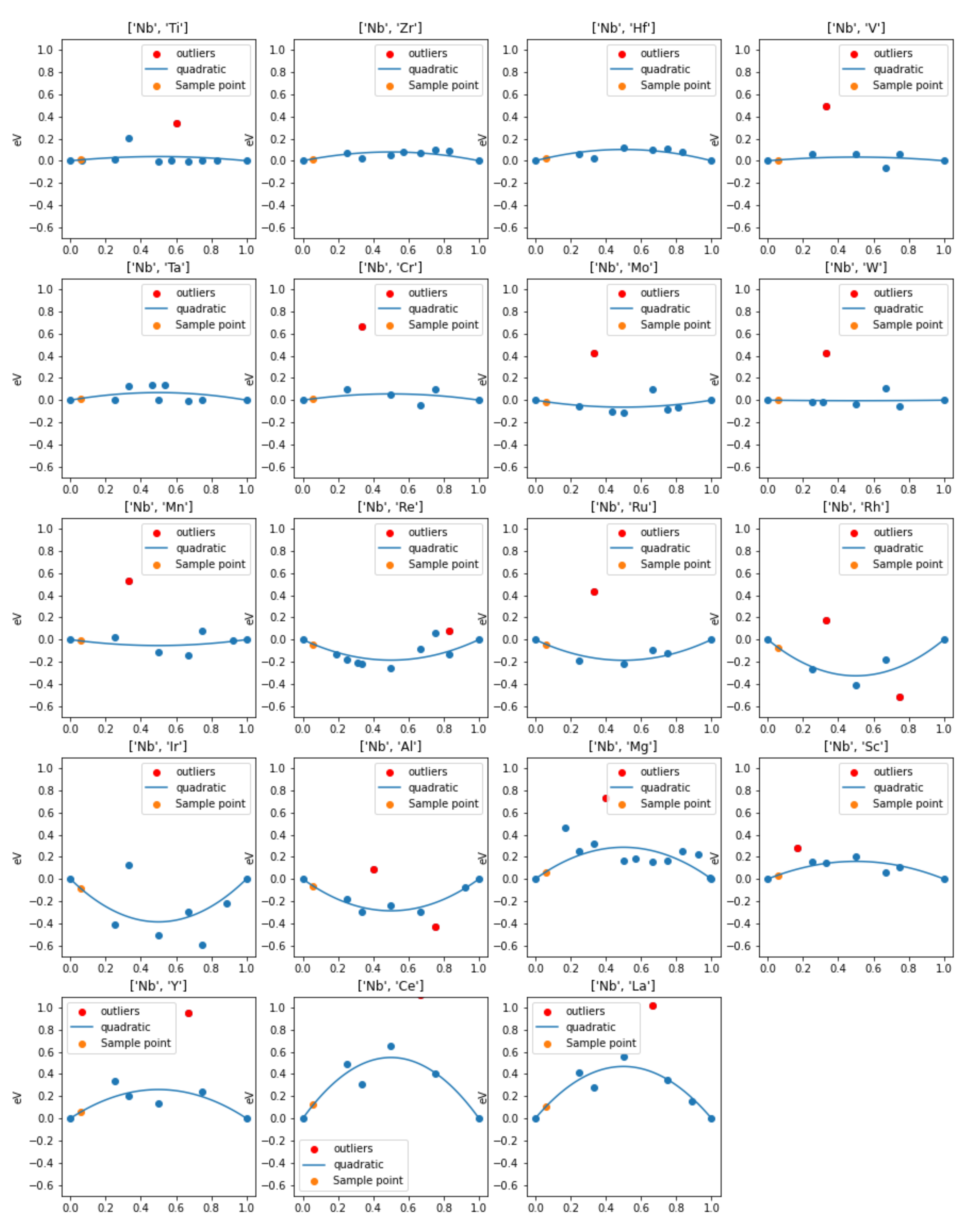}
  \end{center}
  \caption{Binary formation enthalpy
    $\Delta H_{\alpha,\beta}^{\mathrm{alloy}}(x_{\alpha}) =
    A_{\alpha,\beta}x_{\alpha}(1-x_{\alpha})$ fit (blue line) to
    formation energies of ordered intermetallic structures (blue
    points) in the OQMD for various elemental pairs including Nb. The
    abscissa is $1-x_{\mathrm{Nb}}$. Red points are considered
    outliers and not included in the fit. The orange point indicates
    the region of enthalpy values that contributes to the
    multicomponent alloy enthalpy $\Delta H^{\mathrm{alloy}}$ for a
    Nb-rich alloy.}
  \label{fig:additional_binary_fits}
\end{figure}

In Fig. \ref{fig:additional_binary_fits}, we show additional examples
of binary enthalpy fits for elemental pairs including Nb. As discussed
in the main text, the fits capture the overall enthalpy trends
reasonably well.

\newpage

\section{Comparison to Miedema model}

\begin{figure}[htbp]
  \begin{center}
    \includegraphics[width=0.5\linewidth]{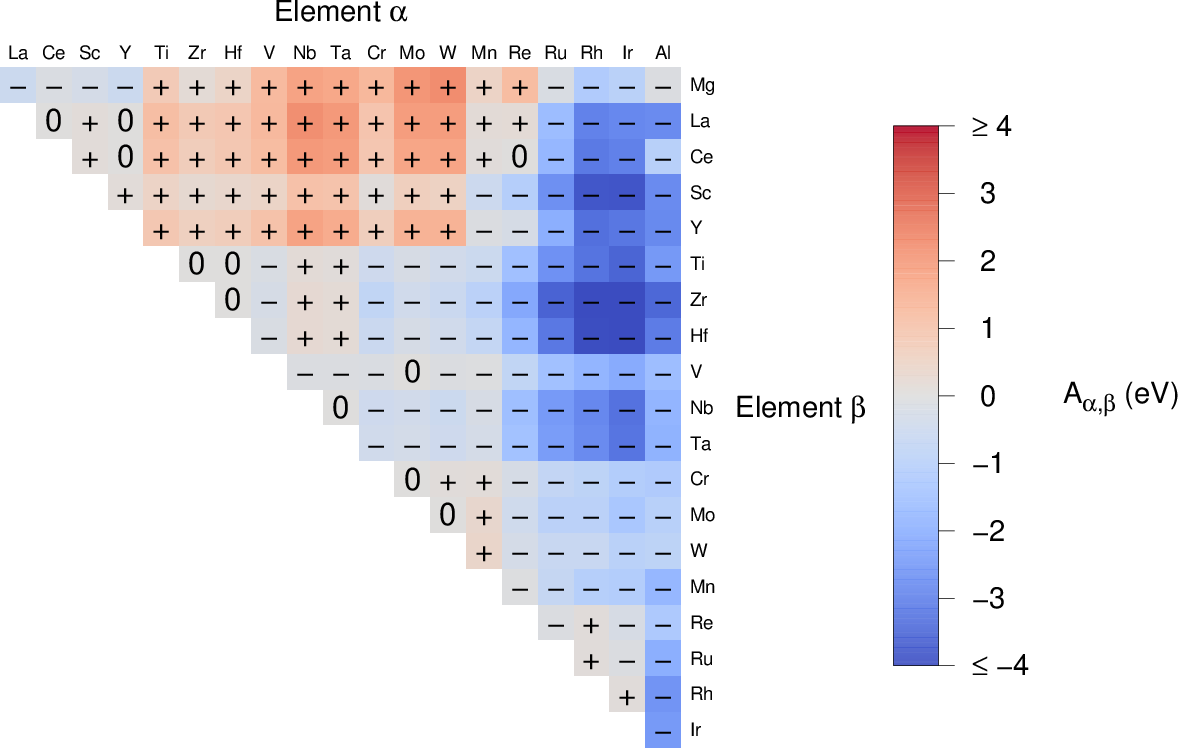}
  \end{center}
  \caption{$A_{\alpha,\beta}$ values from the semiempirical Miedema
    model. Only the top-right of the table is shown as the bottom-left
    contains redundant information
    ($A_{\alpha,\beta}= A_{\beta,\alpha}$).}
  \label{fig:miedema}
\end{figure}

Figure \ref{fig:miedema} shows the $A_{\alpha,\beta}$ parameters
obtained via the semiempirical Miedema model, as implemented in the
\texttt{qmpy} package.\cite{kirklinOpenQuantumMaterials2015} The
values are obtained as four times the Miedema model formation enthalpy
for an equiatomic ($x=1/2$) binary composition, with the factor of
four stemming from the $x(1-x)$ factor in the zeroth-order
Redlich-Kister polynomial form. The Miedema model values exhibit
similar broad trends as the $A_{\alpha,\beta}$ values in our
thermodynamic model, such as (1) the strong region of negative values
for many elemental pairs including Rh or Ir and (2) the region of
substantially positive values for many elemental pairs including La or
Ce.

\newpage

\section{Additional details on experimental validation}

Table \ref{tab:annealing} shows additional information about the
experimental validation. Included are the compositions, whether a
single-phase (S) or multiphase (M) system was observed, the digital
object identifier (DOI) for the literature reference, and the
annealing temperature. The annealing temperatures were obtained
manually from the corresponding papers since this information is not
contained within the database of Borg \textit{et al.} itself.

In assessing the performance of the model of Troparevsky \textit{et
  al.}, we employ (1) an upper formation energy limit of 37 meV/atom
and (2) a lower formation energy limit of $-232$ meV/atom, which is
used in their work in evaluation of refractory alloys (which
characterizes the compositions in Table \ref{tab:annealing}).

\begin{center}
  \begin{longtable}{|l|l|l|r|}
    \caption{The composition, single-phase or multiphase nature,
    literature reference, and annealing temperature for the
    experimental validation of the thermodynamic model developed in
    this work.}\label{tab:annealing}\\\hline
    Composition & Single (S) or Multiple (M) Phases & DOI & Annealing temperature (K)\\
    \hline
    HfNbTaTiZr & S & 10.1007/s10853-012-6260-2 & 1473\\
    AlCrMoNbTi & S & 10.1007/s11661-017-4386-1 & 1473\\
    AlMoNbTi & S & 10.1007/s11661-017-4386-1 & 1773\\
    AlCrMoTi & S & 10.1007/s11661-017-4386-1 & 1573\\
    HfNbTaTiZr & S & 10.1007/s11661-018-4646-8 & 1073\\
    Al0.3NbTa0.8Ti1.4V0.2Zr1.3 & S & 10.1007/s11837-014-1066-0 & 1473\\
    Al0.3NbTaTi1.4Zr1.3 & M & 10.1007/s11837-014-1066-0 & 1473\\
    Al0.4Hf0.6NbTaTiZr & S & 10.1007/s11837-014-1066-0 & 1473\\
    Al0.5NbTa0.8Ti1.5V0.2Zr & M & 10.1007/s11837-014-1066-0 & 1473\\
    AlMo0.5NbTa0.5TiZr & M & 10.1007/s11837-014-1066-0 & 1673\\
    AlNb1.5Ta0.5Ti1.5Zr0.5 & S & 10.1007/s11837-014-1066-0 & 1673\\
    NbZrTi & S & 10.1007/s12598-019-01310-6 & 1373\\
    NbHfZrTi & S & 10.1007/s12598-019-01310-6 & 1373\\
    HfNbTaZr & M & 10.1016/j.actamat.2016.01.018 & 2073\\
    HfNbTaTiZr & M & 10.1016/j.actamat.2017.09.035 & 773\\
    HfNbTaTiZr & S & 10.1016/j.actamat.2017.09.035 & 1373\\
    HfNbTaTiZr & S & 10.1016/j.actamat.2017.09.062 & 1348\\
    MoNbTaTi & S & 10.1016/j.actamat.2019.06.006 & 1673\\
    WNbTaTi & S & 10.1016/j.actamat.2019.06.006 & 1673\\
    HfNbTaTi & S & 10.1016/j.actamat.2019.06.006 & 1673\\
    CrNbMoTi & S & 10.1016/j.actamat.2019.06.006 & 1673\\
    CrMoTaTi & M & 10.1016/j.actamat.2019.06.006 & 1673\\
    NbTaTi & S & 10.1016/j.actamat.2019.06.006 & 1673\\
    AlCrMoNbTi & M & 10.1016/j.actamat.2019.06.032 & 1473\\
    AlMoNbTi & S & 10.1016/j.actamat.2019.06.032 & 1473\\
    NbTiZr & S & 10.1016/j.actamat.2019.06.032 & 1473\\
    MoNbTaW & S & 10.1016/j.actamat.2019.06.032 & 1473\\
    MoNbTi & S & 10.1016/j.actamat.2019.06.032 & 1473\\
    NbTa0.3TiZr & S & 10.1016/j.actamat.2019.06.032 & 1473\\
    Al0.24NbTiVZr & M & 10.1016/j.actamat.2019.06.032 & 1473\\
    Cr0.3NbTiZr & M & 10.1016/j.actamat.2019.06.032 & 1473\\
    NbRe0.3TiZr & M & 10.1016/j.actamat.2019.06.032 & 1473\\
    NbTaTiVZr & M & 10.1016/j.actamat.2019.06.032 & 1473\\
    MoNbTaVW & S & 10.1016/j.actamat.2019.06.032 & 1473\\
    MoNbTiVZr & M & 10.1016/j.actamat.2019.06.032 & 1473\\
    AlCrMoTi & S & 10.1016/j.actamat.2019.06.032 & 1473\\
    HfNbTaTiZr & S & 10.1016/j.actamat.2019.06.032 & 1473\\
    Al0.4Hf0.6NbTaTiZr & S & 10.1016/j.actamat.2019.06.032 & 1473\\
    AlMo0.5NbTa0.5TiZr & M & 10.1016/j.actamat.2019.06.032 & 1473\\
    CrMo0.5NbTa0.5TiZr & M & 10.1016/j.actamat.2019.06.032 & 1473\\
    CrHfNbTiZr & M & 10.1016/j.ijrmhm.2014.07.009 & 1173\\
    HfNbTiVZr & M & 10.1016/j.ijrmhm.2014.07.009 & 1173\\
    NbHfZrTi & S & 10.1016/j.intermet.2019.04.003 & 1563\\
    HfNbTaTiZr & S & 10.1016/j.jallcom.2015.07.209 & 1473\\
    HfNbTaTiZr & M & 10.1016/j.jallcom.2015.07.209 & 1073\\
    AlCrNbTiV & M & 10.1016/j.jallcom.2015.08.224 & 1473\\
    AlCr1.5NbTiV & M & 10.1016/j.jallcom.2015.08.224 & 1473\\
    AlCr0.5NbTiV & S & 10.1016/j.jallcom.2015.08.224 & 1473\\
    AlNbTiV & S & 10.1016/j.jallcom.2015.08.224 & 1473\\
    HfNbTaTiZr & S & 10.1016/j.jallcom.2019.04.291 & 1473\\
    HfNbTaTiZr & S & 10.1016/j.jallcom.2019.05.322 & 1373\\
    HfNbTaTiZr & S & 10.1016/j.matchar.2018.07.015 & 1273\\
    HfMoNbTiZr & S & 10.1016/j.matdes.2015.05.019 & 1743\\
    NbTiV0.3Mo0.3Zr & S & 10.1016/j.matdes.2015.06.072 & 1273\\
    NbTiV0.3Mo0.7Zr & M & 10.1016/j.matdes.2015.06.072 & 1273\\
    AlMo0.5NbTa0.5TiZr0.5 & S & 10.1016/j.matdes.2017.11.033 & 1673\\
    Al0.25NbTaTiZr & M & 10.1016/j.matdes.2017.11.033 & 1673\\
    AlMo0.5NbTa0.5TiZr & M & 10.1016/j.matdes.2017.11.033 & 1673\\
    Al0.5Mo0.5NbTa0.5TiZr & M & 10.1016/j.matdes.2017.11.033 & 1673\\
    AlNbTa0.5TiZr0.5 & S & 10.1016/j.matdes.2017.11.033 & 1673\\
    HfNbTaTi & S & 10.1016/j.matdes.2018.06.003 & 1673\\
    CrMoTaTi & M & 10.1016/j.matdes.2018.06.003 & 1673\\
    AlCrMoNb & M & 10.1016/j.matdes.2018.06.003 & 1673\\
    AlHfNbTi & S & 10.1016/j.matdes.2018.06.003 & 1673\\
    MoNbTaTi & S & 10.1016/j.matdes.2018.06.003 & 1673\\
    AlHfTaTi & S & 10.1016/j.matdes.2018.06.003 & 1673\\
    AlNbTaTi & M & 10.1016/j.matdes.2018.06.003 & 1673\\
    CrNbTaTi & M & 10.1016/j.matdes.2018.06.003 & 1673\\
    AlMoNbTi & M & 10.1016/j.matdes.2018.06.003 & 1673\\
    CrNbTiW & M & 10.1016/j.matdes.2018.06.003 & 1673\\
    AlNbTiV & S & 10.1016/j.matlet.2014.11.162 & 1473\\
    HfNbTaTiZr & S & 10.1016/j.matlet.2016.08.060 & 1523\\
    Al0.5CrNbTi2V0.5 & M & 10.1016/j.matlet.2016.11.030 & 1473\\
    TiNbTa0.5ZrAl0.5 & M & 10.1016/j.matlet.2019.03.075 & 1273\\
    CrMo0.5NbTa0.5TiZr & M & 10.1016/j.msea.2011.09.033 & 1723\\
    CrNbTiZr & M & 10.1016/j.msea.2012.12.018 & 1473\\
    CrNbTiVZr & M & 10.1016/j.msea.2012.12.018 & 1473\\
    NbTiVZr & S & 10.1016/j.msea.2012.12.018 & 1473\\
    NbTiV2Zr & M & 10.1016/j.msea.2012.12.018 & 1473\\
    AlNbTiVZr0.1 & M & 10.1016/j.msea.2017.08.019 & 1473\\
    AlNbTiVZr0.25 & M & 10.1016/j.msea.2017.08.019 & 1473\\
    AlNbTiVZr0.5 & M & 10.1016/j.msea.2017.08.019 & 1473\\
    AlNbTiVZr1.5 & M & 10.1016/j.msea.2017.08.019 & 1473\\
    AlNbTiVZr & M & 10.1016/j.msea.2017.08.019 & 1473\\
    AlNbTiV & S & 10.1016/j.msea.2017.08.019 & 1473\\
    Hf0.75NbTa0.5Ti1.5Zr1.25 & S & 10.1016/j.msea.2017.10.073 & 1270\\
    MoNbTaTiZr & M & 10.1016/j.scriptamat.2016.10.028 & 1273\\
    AlNbTiZr & M & 10.1080/02670836.2018.1446267 & 1473\\
    HfNb0.18Ta0.18Ti1.27Zr & S & 10.1080/21663831.2016.1221861 & 1123\\
    MoNbTaV & S & 10.3390/e18050189 & 1673\\\hline
  \end{longtable}
\end{center}

\bibliography{stability_model}